\documentclass[sort&compress,final,3p,times,twocolumn,fixfloat]{elsarticle}
\usepackage{comment}
\usepackage{graphicx,color}
\usepackage{lineno}
\journal{Physics Letters B}

\newcommand{\eqs}[1]{\begin{equation} \begin{split} #1\end{split} \end{equation} }
\newcommand{\nn}{\nonumber}

\newcommand{\ce}[1]{Eq.~(\ref{#1})}
\newcommand{\cf}[1]{{Fig.~\ref{#1}}}

\usepackage{xfrac}
\usepackage{txfonts}
\usepackage{bm}
\usepackage{amsmath}
\usepackage{amssymb}
\usepackage{booktabs}
\usepackage{float}

\usepackage{xspace} 
\usepackage{mathtools, cuted}

\usepackage{scalerel}
\usepackage{orcidlink}
\usepackage{tikz}
\usepackage{slashed}
\usetikzlibrary{svg.path}

\DeclareMathAlphabet{\pazocal}{OMS}{zplm}{m}{n}
\newcommand{\Q}{\pazocal{Q}}

\usepackage[normalem]{ulem}

\newcommand{\new}[1]{{\color{blue}#1}}

\newcommand*\oline[1]{%
   \vbox{%
     \hrule height 0.5pt%
     \kern0.4ex%
     \hbox{%
       \kern-0.15em%
       \ifmmode#1\else\ensuremath{#1}\fi%
       \kern-0.15em%
     }%
   }%
}

 \usepackage{ifpdf}
 \ifpdf
 \usepackage[]{hyperref}
 \else
 \usepackage[hypertex]{hyperref}
 \fi

 \hypersetup{
   pdftitle={},%
   pdfauthor={},%
   pdfsubject={},%
   pdfkeywords={},%
   pdfstartview={},%
   bookmarksopen=true, breaklinks=true, debug=true, %
   colorlinks=true, linkcolor=red, citecolor=blue, urlcolor=blue
 }

\usepackage{float} %
\usepackage[caption=false]{subfig} %

\usepackage[sort&compress, capitalise]{cleveref}

\newcount\savefnused
\newcount\savefndone

\newcommand{\savefootnote}[2][\empty]%
{\ifx\empty#1\footnotemark\else\footnotemark[#1]\fi
 \global\advance\savefnused by 1
 \expandafter\xdef\csname savefnmark\the\savefnused\endcsname{\thefootnote}%
 \expandafter\xdef\csname savefntext\the\savefnused\endcsname{#2}%
}
\newcommand{\flushfootnote}{\loop\ifnum\savefndone<\savefnused
  \global\advance\savefndone by 1
  \footnotetext[\csname savefnmark\the\savefndone\endcsname]%
    {\csname savefntext\the\savefndone\endcsname}%
  \global\expandafter\let\csname savefnmark\the\savefndone\endcsname\relax
  \global\expandafter\let\csname savefntext\the\savefndone\endcsname\relax
\repeat}

\graphicspath{ {} }

\begin{document}

\title{Exclusive vector-quarkonium photoproduction at NLO in $\alpha_s$ in collinear factorisation with evolution of the generalised parton distributions and high-energy resummation}

\date{\today}

\address[IJCLab]{Universit\'e Paris-Saclay, CNRS, IJCLab, 91405 Orsay, France}
\address[NCBJ]{National Centre for Nuclear Research (NCBJ), Pasteura 7, 02-093 Warsaw, Poland}

\author[IJCLab]{C.A. Flett\,\orcidlink{0000-0002-6295-3793}}
\ead{Christopher.Flett@ijclab.in2p3.fr}
\author[IJCLab]{J.P. Lansberg\,\orcidlink{0000-0003-2746-5986}}
\ead{Jean-Philippe.Lansberg@ijclab.in2p3.fr}
\author[IJCLab]{S. Nabeebaccus\,\orcidlink{0000-0003-1842-7929}}
\ead{Saad.Nabeebaccus@ijclab.in2p3.fr}
\author[IJCLab]{M. Nefedov\,\orcidlink{0000-0002-1046-9625}}
\ead{Maxim.Nefedov@ijclab.in2p3.fr}
\author[NCBJ]{P. Sznajder\,\orcidlink{0000-0002-2684-803X}}
\ead{Pawel.Sznajder@ncbj.gov.pl}
\author[NCBJ]{J. Wagner\,\orcidlink{0000-0001-8335-7096}}
\ead{Jakub.Wagner@ncbj.gov.pl}

\begin{abstract}
We perform the first complete one-loop study of exclusive photoproduction of vector quarkonia off protons in Collinear Factorisation (CF) including the scale evolution of the Generalised Parton Distributions (GPDs). We confirm the perturbative instability of the cross section at high photon-proton-collision energies 
($W_{\gamma p}$)
at Next-to-Leading Order (NLO) in $\alpha_s$ and solve this issue by resumming higher-order QCD corrections, %
which are enhanced by a logarithm of the parton energies, using High-Energy Factorisation (HEF) in the {Doubly-Logarithmic Approximation (DLA)} matched to CF. %
Our NLO CF $\oplus$ DLA HEF results are in agreement with the latest HERA data, show a smaller sensitivity to the factorisation and renormalisation scales compared to Born-order results. Quark-induced channels  via interference with gluon ones are found to contribute at most 20\% of the cross section for $W_{\gamma p} > 100$~GeV. %
Our results also show that such exclusive cross sections cannot be accurately obtained from the square of usual Parton Distribution Functions (PDFs) 
and clearly illustrate the importance of quarkonium exclusive photoproduction to advance our understanding of the 3D content of the nucleon in terms of gluons.
Our work provides an important step towards a correct interpretation of present and future experimental data collected at HERA, the EIC, the LHC and future experiments.
\end{abstract}

\begin{keyword}
$J/\psi$, $\Upsilon$, exclusive photoproduction, HERA, EIC, LHC, LHeC, FCC-eh, EIcC,  NLO, high-energy resummation
\end{keyword}
\maketitle

\section{Introduction}\label{sec:introduction}
Hard exclusive reactions 
are complementary to inclusive reactions such as Deep-Inelastic Scattering (DIS) in order to probe the inner content of the hadrons and study the strong interaction.
A class of such hard exclusive reactions is of particular interest 
as they probe the 3D hadron structure through Generalised Parton Distributions~(GPDs) \cite{Radyushkin:1997ki,Burkardt:2000za,Burkardt:2002hr}
which extend the usual 1D Parton Distribution Functions (PDFs) 
of Collinear Factorisation (CF) used 
for single-scale hard-inclusive reactions like DIS.

\begin{figure}[!hbt]
\centering
\includegraphics[width=0.8\columnwidth,angle =0]{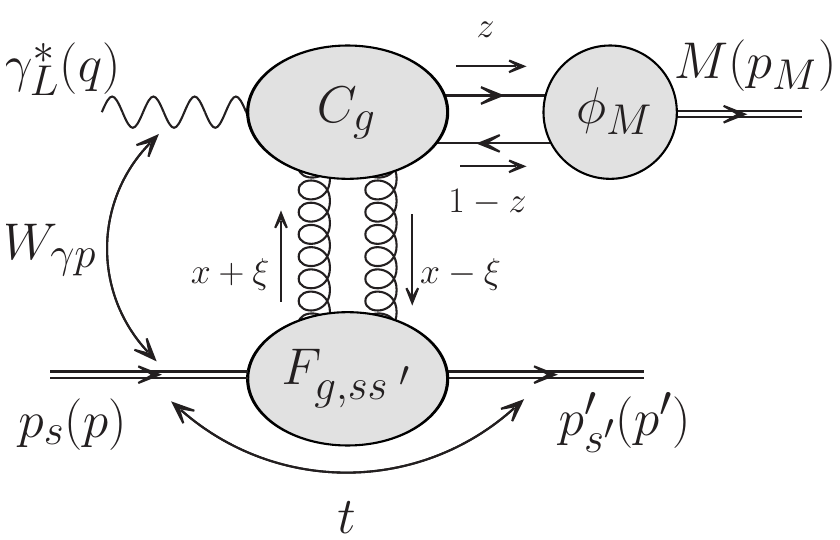}\vspace*{-.2cm}
\caption{Factorisation of the amplitude for DVMP in the gluon channel.}
\label{fig:DVMP}
\end{figure}

One such process is exclusive Deeply-Virtual-Meson-Production (DVMP) off protons, initiated by a longitudinally polarised photon, $\gamma_L^*(q) p_s(p) \rightarrow M(p_M) p'_{s'}(p')$, see~\cf{fig:DVMP}. 
This process admits, at small $t=(p'-p)^2$ and in the massless-quark limit, a collinear factorisation~\cite{Collins:1996fb} at the level of the amplitude, ${\cal A}_{ss'}$, as a double convolution of (i) a process-dependent perturbative hard-scattering coefficient function, $C_i$, (ii) a universal, non-perturbative meson distribution amplitude (DA), $\phi$, and (iii) a {universal, non-perturbative} quark and/or gluon GPD correlator ${F}_{i,ss'}$ $(i=q,g)$
, as
 \begin{equation}
    {\cal A}_{ss'}= {F}_{i,ss'}(x,\xi,t;\mu_F)  \otimes C_i(x,\xi,z;\mu_F,\mu_R) \otimes \phi_M(z;\mu_F).
    \label{eq:amplitude-factorisation}
\end{equation}
  Here, $\otimes$ denotes the sum of the flavour, spin and colour of the exchanged partons, and the convolution over $x$ and $z$, with $z$ being the fraction of the meson $M$ momentum carried by the quark, and $x$ being the average momentum {fraction} of the nucleon $p$ carried by the parton $i$. In addition, $\Delta= (p'-p)$ is the momentum transfer and\footnote{Here, the light-cone components are defined as $k^{\pm}\equiv(k^0\pm k^3)/\sqrt{2}$ and $\mathbf{k}_\perp\equiv (k^1,k^2)$ in the $\gamma p$ center-of-momentum frame, with the $z$ axis defined along the proton direction.%
 } $\xi = \Delta^+/2P^+$ ($P=(p+p')/2$) is the fraction of the longitudinal proton momentum transfer, or skewness. The presence of $x$, $\xi$ and $t$ in ${F}$ underlies the factorisation in terms of 3D GPDs rather than the usual 1D PDFs and the presence of $\mu_F$ is due to the fact that both GPDs and DAs are subject to evolution equations.

In the light-cone gauge and at leading twist,  ${F}_q$ and ${F}_g$ are Fourier transform\new{s} of  matrix elements of chiral-even operators constructed from quark fields $\psi^q$ or gluon-field-strength tensors $F^{\mu \, \nu}$ as follows (with $y^+ = 0$ and  $\mathbf{y}_{\perp} = \mathbf{0}$):
\eqs{
    &{F}_{q,ss'} = \frac{1}{2} \! \int \! \frac{\text{d} y^-}{2 \pi} e^{i x P^+ y^-} \langle p',s' | \bar{\psi}^q \left(\tfrac{-y}{2} \right) \gamma^+ \psi^q \left(\tfrac{y}{2} \right) | p,s \rangle, \\
    &{F}_{g,ss'} = \frac{1}{P^+} \! \int\! \frac{\text{d} y^-}{2 \pi} e^{i x P^+ y^-} \langle p',s' | F^{+ \mu}\left(\tfrac{-y}{2} \right) F^{~+}_{\mu} \left(\tfrac{y}{2} \right) | p,s \rangle . 
}

In practice, they are parametrised  via specific Lorentz structures and twist-2 parton-helicity conserving GPDs $H^{q,g}(x,\xi,t;\mu_F)$ and $E^{q,g}(x,\xi,t;\mu_F)$,  as follows ($j=q,g$)%
\eqs{
    &{F}_{j,ss'} 
    =\frac{1}{2 P^+} \left[ \bar{u}_{s'}(p') \left(H_j   \gamma^+  + E_j  \frac{ i \sigma^{+ \Delta}}{2m_p}\right) u_s(p) \right].
    \label{GPD-parametrisation}
}

Although not formally proven to all orders in perturbative QCD~(pQCD), the same factorisation theorem is assumed for photoproduction (i.e. quasi-real photons with virtuality $q^2 \approx 0$) of heavy vector quarkonia, $\gamma p \rightarrow \Q p$, where $\Q = J/\psi, \Upsilon, \dots$, due to the hard scale provided by the heavy-quark mass $m_{Q}$. Explicit one-loop computations in the non-relativistic static limit, where the DA $\phi_{M}(z;\mu_F)$ reduces to $\delta(z-{1}/{2})$, have shown that factorisation holds up to NLO in $\alpha_s$~\cite{Ivanov:2004vd}. This also applies for off-shell photons~\cite{Flett:2021ghh}.
While the LO cross section uniquely depends on gluon GPDs, sensitivity to quark GPDs arises at NLO along with  explicit renormalisation ($\mu_R$) and factorisation ($\mu_F$) scale dependences. %

However, the corresponding NLO cross sections are very sensitive to $\mu_F$ for $W_{\gamma p} \gg m_Q$, like for several inclusive quarkonium observables~\cite{Schuler:1994hy,Mangano:1996kg,Feng:2015cba}. In these inclusive reactions, this has been resolved either by fixing  $\mu_F$ to reduce anomalously large NLO corrections~\cite{Lansberg:2020ejc,Serri:2021fhn} or by resumming the high-energy logarithms responsible for these perturbative instabilities using High-Energy Factorisation (HEF)~\cite{Lansberg:2021vie,Lansberg:2023kzf}. 
Here, we aim  to demonstrate that a similar resummation also cures the issue in exclusive reactions,  going beyond the scale-fixing criterion advocated in~\cite{Jones:2015nna} and to perform a complete phenomenological study with GPD evolution, including comparisons to precise HERA data.%

The structure of this Letter is as follows: Section 2 reviews CF formulae at fixed order in $\alpha_s$, identifies the origin of the perturbative instabilities at NLO, and explains our GPD modelling and evolution set up. Section 3 explains HEF resummation and the limitations of a scale-fixing criterion beyond NLO. Section 4 compares our results to experimental data and analyses resummation effects on scale dependencies. Section 5 discusses connections with other approaches using forward limits.
Section 6 gathers our conclusions.

\section{Exclusive photoproduction of vector quarkonia in collinear factorisation}
\subsection{GPD modelling and evolution}
\label{sec:GPD-evo}

A very convenient procedure to construct models of GPDs encapsulating all their properties is to use double distributions~\cite{Radyushkin:1997ki}, $f_{i}(\beta,\alpha)$, where $i=q,g$. These distributions are related to GPDs in the following way:
\eqs{
H_{i}(x,\xi;\mu_0) \! = \!\int_{-1}^{1}   \!\!\!d\beta \! \int^{1-|\beta|}_{-1+|\beta|}d\alpha  \delta(\beta+\xi\alpha-x) f_{i}(\beta,\alpha;\mu_0).
}
Assuming that $f_{i}(\beta,\alpha)$ are even functions of $\alpha$, polynomiality of GPDs, which is a consequence of Lorentz invariance (see e.g. Ref.~\cite{Ji:1998pc}), is automatically satisfied. 

One can further ensure the proper forward limits\footnote{$H_g(\pm x,0;\mu_F)=xg(x;\mu_F)$ and $H_q(x,0;\mu_F)=q(x;\mu_F)$, $H_q(-x,0;\mu_F)=-\bar{q}(x;\mu_F)$ for $x > 0$.} of the GPDs, connecting them to PDFs, by the following common factorisation Ansatz~\cite{Goeke:2001tz,Belitsky:2001ns}: 
\begin{equation}
f_{i}(\beta,\alpha;\mu_0) = h_{i}(\beta,\alpha)\times \left\{ \begin{array}{cc}
    |\beta| g(|\beta|;\mu_0) & \text{for }i=g,  \\
    \theta(\beta) q_{\rm val}(|\beta|;\mu_0) & \text{for valence }q, \\
    {\rm sgn}(\beta) q_{\rm sea}(|\beta|;\mu_0) & \text{for sea }q,
\end{array} \right. \label{eq:dd-model}
\end{equation}
where $g(x)$ is the gluon PDF, $q_{\mathrm{val}}(x)$ and $q_{\mathrm{sea}}(x)$ are the valence and sea components of quark PDFs and $h_{i}(\beta,\alpha)$ is the so-called profile function satisfying
\begin{equation}
    \int_{-1+|\beta|}^{1-|\beta|} d\alpha\, h_{i}(\beta,\alpha)  = 1,
\end{equation}
as originally proposed in~\cite{Radyushkin:1998es,Radyushkin:1998bz}.
A popular choice for $h_{i}(\beta,\alpha)$ is then~\cite{Musatov:1999xp}: 
\begin{equation}h_i(\beta, \alpha) \equiv \frac{\Gamma(2n_{i}+2)}{2^{2n_{i}+1}\Gamma^{2}(n_{i}+1)}
\frac{\left( (1-|\beta|)^2 - \alpha^2\right)^{n_{i}}}{(1-|\beta|)^{2n_{i}+1}}\,,
\end{equation}
where the parameter $n_i$ (which can be flavour-dependent) controls the width of the profile function and, effectively, the buildup of the skewness effect. In particular, for $n \to \infty$, the $\xi$ dependence of such a GPD disappears, i.e. ${h}_q(\beta,\alpha)|_{n \to \infty}=\delta(\alpha)$ and therefore $H_q{(x,\xi;\mu_0)|_{n \to \infty}}=q(x;\mu_0)$.

Along the same lines, a $t$ dependence can easily be implemented, while keeping the same limiting behaviours, by extending   $h_i(\beta,\alpha)$  to $h_i(\beta, \alpha, t)$ satisfying $\lim_{t\rightarrow 0} h_{i}(\beta,\alpha,t) = h_{i}(\beta,\alpha)$.
In the Goloskokov-Kroll~(GK) DD model~\cite{Goloskokov:2006hr}, which we use here, it is done via a simple exponential dependence in $t$ such that $h_i(\beta, \alpha, t) \equiv e^{b t} h_i(\beta,\alpha)$ while $n_g = n_q^{\mathrm{sea}} = 2$ and $n_q^{\mathrm{val}} = 1$.
Instead of using the CTEQ6M PDF set~\cite{Pumplin:2002vw} as the forward inputs as in~\cite{Goloskokov:2006hr}, one can use more modern PDF sets. For most of our results, we will use the central set of CT18NLO~\cite{Hou:2019efy}.  As we explain in the next section, we focus on the differential cross section at the minimum value of $|t|$, $t_\text{min}=-4\xi^2 m_p^2/(1-\xi^2)$ which, at large $W_{\gamma p}$ where $\xi \simeq M^2_\Q/(2W_{\gamma p}^2)$, becomes $t_\text{min} \simeq -m_p^2 M^4_\Q/W_{\gamma p}^4$ and is thus extremely small compared to the other hadronic scales. As such, we do not need to model the $t$ dependence of GPDs. %

This DD model has been implemented in the \texttt{PARTONS} framework~\cite{Berthou:2015oaw} interfaced to the \texttt{APFEL++} code~\cite{Bertone:2017gds} for performing the full leading-logarithmic~(LL) GPD evolution with kernels at one-loop~\cite{Bertone:2022frx} and has been used at $\mu_0=2~\mathrm{GeV}$ as our initial condition for the GPD evolution.

We note that modelling GPDs with DDs only allows one to study a class of GPD models. In particular, they do not allow one to include
contributions with the largest $\xi$ power  which are usually modelled by the so-called $D$ term~\cite{Polyakov:1999gs}.   At large $W_{\gamma p}$ (thus small $\xi$) which is our focus here, these are, however, not expected to be relevant.

\subsection{Coefficient functions and cross sections in Collinear Factorisation}
\label{sec:C-NLO}
Following Ref.~\cite{Flett:2021ghh}, we write the quarkonium photoproduction amplitude in CF at leading twist in the form:%
\eqs{ \!\! {\mathcal A_{{ss'}}^{\lambda \lambda'}}=-\varepsilon_{\lambda}^{\gamma}.\varepsilon_{\lambda'}^{*\Q}\! \sum\limits_{i={q,g}} \int\limits_{-1}^1\!\! \frac{dx}{x^{1+\delta_{ig}}}  C_{ i}\left(x,\xi;\mu_F, \mu_R\right){F}_{i,ss'}(x,\xi,t;\mu_F),\! \label{eq:Ampl-CF}
}
where $\varepsilon_{\lambda}^{\gamma}$ is the  photon-polarisation vector with helicity $\lambda$ in the gauge $\varepsilon_{\lambda}^{\gamma}.p = 0$ and
$\varepsilon_{\lambda'}^{\Q}$ that of the quarkonium. %
In~\ce{eq:Ampl-CF}, the integration over $x \in [-1,1]$ covers the so-called DGLAP region (with $\xi<|x|<1$) and the ERBL region (with $|x|<\xi<1$), two distinct regions  where the GPDs obey two different evolution equations~\cite{Bertone:2022frx}. %

We first note that the $t$ dependence of the CF amplitude only explicitly appears through the GPD $F_i$. Since the gluon GPDs giving the largest contribution to quarkonium photoproduction are poorly known and their $t$ dependence essentially unknown, we will limit ourselves to $t=t_\text{min}$ which can in practice be approximated to zero for the $W_{\gamma p}$ values attainable at the EIC, HERA and the LHC in Ultra-Peripheral Collisions (UPCs). Precise HERA data~\cite{H1:2013okq} for $t\simeq 0$ as a function of $W_{\gamma p}$ are available and are sufficient for the data-theory comparison we aim at. We leave discussions on the $t$ dependence and of the $t$-integrated cross sections for a future work\footnote{
The $\gamma p \rightarrow \Q p$ two-body $t$-differential cross section is readily obtained from the amplitude~\ce{eq:Ampl-CF} as usual from 
${d \sigma}/{dt} = 1/(16 \pi (W_{\gamma p}^2 - m_p^2)^2)     \overline{\sum}_{\lambda,\lambda'={\pm}}\,\overline{\sum}_{s,s'} |\mathcal A_{ss'}^{\lambda \lambda'}|^2$,
where the sums include the initial-state averaging over the incoming proton spins $(s,s')$ and the photon transverse helicities $(\pm)$. 
$\sigma$ is then obtained by integrating from $t_\text{min}$ to $-\infty$.}.

As we announced, at LO in $\alpha_s$ and in the squared relative velocity of the heavy quarks, $v^2$, the gluon coefficient function (unlike the quark one) is nonzero:%
\begin{equation}
 C_{g}^{\text{(0)}}\left(x,\xi;\mu_R \right) =  \frac{x^2 c}{(x + \xi -i\varepsilon)(x - \xi +i\varepsilon ) },  \label{eq:Cg-LO}
\end{equation}
where $c=(4\pi \alpha_s(\mu_R) e e_{Q} R_\Q(0))/(m_Q^{3/2}\sqrt{2\pi N_c})$.  In the above, $R_\Q(0)$ is the value at the origin of
the spatial radial wave function\footnote{It is the solution of the non-relativistic Schr\"odinger equation.} of the quarkonium $\Q$, 
$e_Q$ is the electric charge of the heavy quark $Q$ in units of the positron charge and $e$ is the positron charge in natural units.
The heavy-quark mass $m_Q$ is chosen to be $M_{\Q}/2$ and we choose $R_{J/\psi}(0) = 1~{\rm GeV}^{3/2}$ and $R_{\Upsilon}(0) = 3~{\rm GeV}^{3/2}$
in the range of potential-model values~\cite{Eichten:1995ch,Eichten:2019hbb}. A two-loop \texttt{CRunDec}~\cite{Schmidt:2012az} running of $\alpha_s$ with $\alpha_s(M_Z^2) = 0.118$ is employed.

At small $\xi$ (or large $W_{\gamma p}$) and finite $x$, the NLO coefficient functions $C^{\text{(1)}}_{g,q}(x,\xi;\mu_R,\mu_R)$ first obtained in \cite{Ivanov:2004vd} scale like
\eqs{
-\frac{i\pi c |x|}{2 \xi} \frac{{\alpha}_s(\mu_R)}{\pi} \ln \left( \frac{m_{Q}^2}{\mu_F^2} \right)
 \Bigg\{C_A,2C_F\Bigg\}\equiv C_{\{g,q\}}^{\text{(1, asy.)}}(x,\xi;\mu_R,\mu_R). \label{eq:Cg-NLO:low-X} 
}

\begin{figure}[hbt!]
    \centering
                 \captionsetup[subfloat]{captionskip=-0.5cm,margin={0.9\columnwidth,0cm}}
    \subfloat[]{\includegraphics[width=\columnwidth]{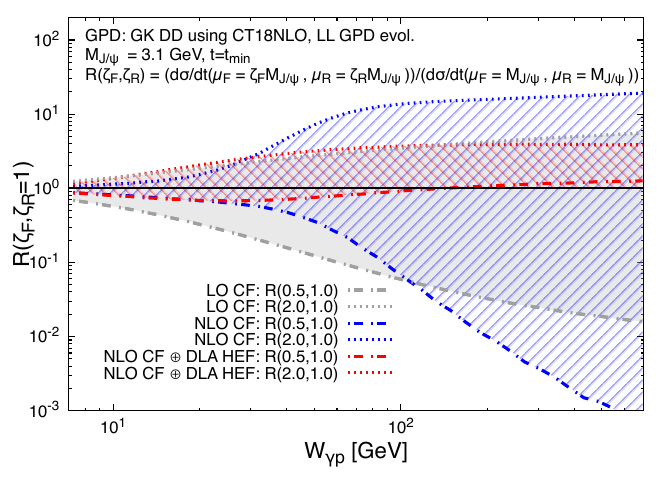} \label{fig:muF-ratio}}\\\vspace*{-.5cm}
   \subfloat[]{ \includegraphics[width=\columnwidth]{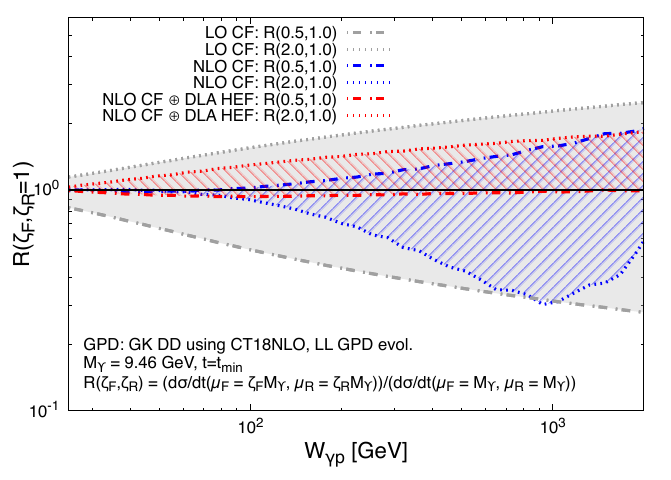}\label{fig:muF-ratio-upsilon}}\vspace*{-.5cm}
        \caption{$\mu_F$ scale uncertainty of the LO CF, NLO CF and NLO CF $\oplus$ DLA HEF (a) $J/\psi$ and (b) $\Upsilon$ $t$-differential  cross sections at $t_\text{min}$.%
    \label{fig:scale_var}}
\end{figure}

Since the $x$ dependence of the gluon GPD ${F}_{g}$ is close to a constant for $\mu_F \sim 2-3$~GeV relevant for $J/\psi$, the corresponding NLO contribution\footnote{A similar reasoning holds for the (NLO) quark contributions with the flavour-singlet GPD scaling like $1/x$.} to the amplitude (\ref{eq:Ampl-CF}) is enhanced by $\ln(1/\xi)$ at small $\xi$,
as observed already in Ref.~\cite{Ivanov:2004vd}. This leads to the  catastrophically large $\mu_F$ dependence shown in~\cf{fig:muF-ratio} of the NLO CF (blue) $J/\psi$ cross section  compared to the LO (grey) one at large $W_{\gamma p}$. The same observation can hardly be made for the $\Upsilon$ case (\cf{fig:muF-ratio-upsilon}) which sits at a larger scale leading to $F_g$ decreasing with $\xi$ and to a smaller $\alpha_s$. Note that the LO CF uncertainty increases with $W_{\gamma p}$ due to GPD evolution (in particular the singular behaviour of $P_{gg}$).

For $\mu_F = m_Q$,  $C_{q,g}^{\text{(1, asy.)}}=0$.
Such a scale choice (denoted $\hat{\mu}_F$ here) corresponds~\cite{Jones:2015nna} to the resummation of a series of corrections to the amplitude $\propto (\alpha_s \ln (\mu_F/m_Q) \ln(1/\xi))^n$, used in phenomenological analyses~\cite{Flett:2019pux,Flett:2020duk,Flett:2022ues}, but misses all corrections $\propto (\alpha_s \ln(1/\xi))^n$ without $\ln(\mu_F)$. A similar scale-fixing prescription was advocated for $P_T$-integrated cross sections of {\it inclusive} hadro-~\cite{Lansberg:2020ejc} and photoproduction~\cite{Serri:2021fhn} of heavy quarkonia. As noted in~\cite{Lansberg:2020ejc}, certain conventional gluon PDFs exhibit a local minimum \new{in $x$} close to $10^{-3}$ for $\mu_F$ values below 2 GeV. In $DD$ models\footnote{The same would apply to the case of GPDs obtained with the Shuvaev transform~\cite{Shuvaev:1999ce,   Shuvaev:1999fm,Dutrieux:2023qnz}.}, this behaviour leads to oscillating GPDs in $x$, via the forward-limit constraints, and results in an unusual energy-dependence of the exclusive $J/\psi$ photoproduction cross section. We however leave this issue to future investigation as it does not relate to the hard-scattering properties but rather to the GPD modelling.

\section{Resummation via High-Energy Factorisation}\label{sec:HEF}

\subsection{The resummed CF coefficient functions}
\label{sec:HEF-coeff-func}

At higher orders in $\alpha_s$, the CF coefficient function,  $C_{ i}\left(x,\xi;\mu_F,\mu_R\right)$, written in terms of the variable $\rho=\xi/x$, develops~\cite{Ivanov:2004vd,Ivanov:2007je} a series of corrections $\propto (\alpha_s^n \ln^{n-1}(1/|\rho|))/|\rho|$ which become important at $|\rho|\ll 1$ and whose consideration leads to the Leading-Logarithmic Approximation (LLA). This series can be resummed using the {\it High-Energy Factorisation} formalism of Refs.~\cite{Catani:1990xk,Catani:1990eg,Collins:1991ty,Catani:1994sq}, originally developed for {\it inclusive} processes. %
The application of this formalism to {\it exclusive} processes is possible due to the %
optical theorem, as explained in the next paragraph.

\begin{figure}[hbt!]
    \centering
    \includegraphics[width=0.32\textwidth]{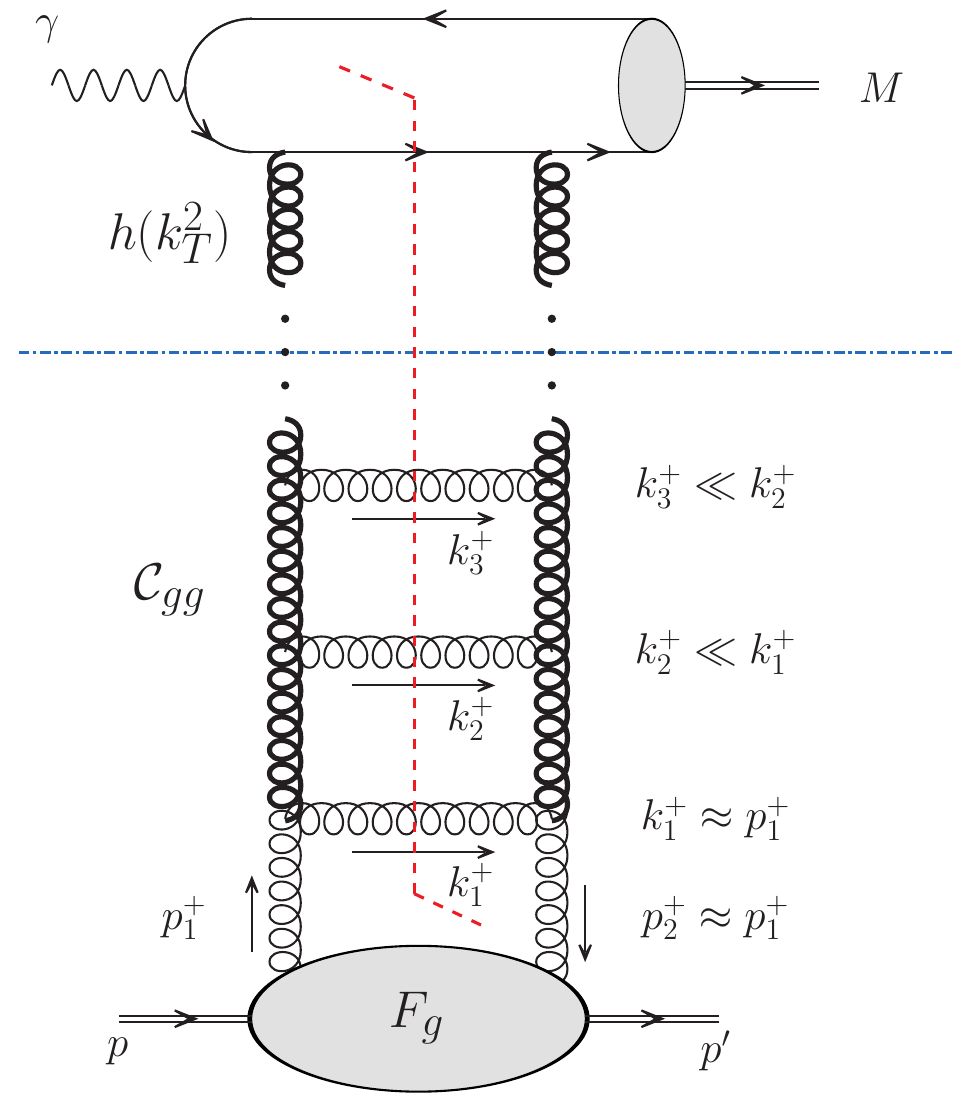}\vspace*{-0.3cm}
    \caption{Typical Feynman diagram contributing to the imaginary part of the photoproduction amplitude (\ref{eq:Ampl-CF}) in the $\rho\ll 1$ region in the LLA. The thick gluons in the $t$ channel are Reggeised. Further notation is explained in the text.
    }
    \label{fig:BFKL}
\end{figure}

The typical Feynman diagram and loop-momentum region contributing to the imaginary part in the LLA is shown in Fig.~\ref{fig:BFKL}, where the gluons depicted by thick helices, connecting emissions strongly ordered in $k^+$, are the so-called {\it Reggeised gluons}; these are the effective gauge-invariant degrees of freedom of high-energy QCD %
whose operator definition can be given e.g. by the Lipatov high-energy EFT~\cite{Lipatov95} or the approach of Ref.~\cite{Caron-Huot:2013fea}, which is equivalent up to NLLA.  One notes that the part of the cut diagram below the horizontal dash-dotted line in Fig.~\ref{fig:BFKL} is just the usual inclusive BFKL ladder~\cite{BFKL1,BFKL2,BFKL3} which is integrated over the transverse momenta of all emitted gluons, including the most energetic %
emission $k_1$. %
Collinear divergences from the ${\bf k}_{i\perp}$ integrations  (\cf{fig:BFKL}) should be subtracted according to the $\overline{\rm MS}$ scheme to comply with the collinear GPD definition. %
This challenging computation~\cite{Catani:1990xk,Catani:1990eg,Catani:1994sq} %
corresponds to the part of the diagram in Fig.~\ref{fig:BFKL} below the horizontal dash-dotted line, i.e. the HEF resummation factor ${\cal C}_{gi}(\rho,{\bf k}_\perp^2;\mu_F,\mu_R)$. 
 This factor has been proven to be process-independent within the LLA~\cite{ReggeProofLLA} and the NLLA~\cite{ReggeProofNLLA} in $\ln(1/|\rho|)$. 
 In the quark channel, one has within the LLA,%
\eqs{
{\cal C}_{gq}=\frac{2C_F}{C_A}\left[  {\cal C}_{gg} %
- \delta(\rho-1)\delta({\bf k}_\perp^2)\right],  \label{eq:C-gq}
}
where the overall factor of $2$ is confirmed by a detailed computation in~\ref{sec:app-A}.

Indeed, the process-dependent HEF coefficient function $h({\bf k}_\perp^2)$, depicted diagrammatically above the dash-dotted line in~\cf{fig:BFKL}, does not contain any large logarithms of $\ln(1/|\rho|)$. As a result, at leading power in $|\rho|\ll 1$, the resummed coefficient function, which is meant to replace $C_i$ in \ce{eq:Ampl-CF}, factorises as follows~\cite{Ivanov:2007je,Ivanov:2015hca}:
\begin{equation}
   \hspace{-2mm} C^{\text{(HEF)}}_{i}(\rho;\mu_F,\mu_R)= \frac{-i}{|\rho|}  \int\limits_0^{\infty} d{\bf k}_\perp^2\ {\cal C}_{gi}(|\rho|,{\bf k}_\perp^2;\mu_F,\mu_R) h({\bf k}_\perp^2),\label{eq:Master-formula-HEF}
\end{equation}
with the well known (see e.g.~\cite{Ivanov:2007je,Jones:2013pga}) HEF coefficient function at LO in $\alpha_s$:%
\begin{equation}
    h({\bf k}_\perp^2) = \frac{\pi c}{2}  \frac{m_{Q}^2}{m_{Q}^2+{\bf k}_\perp^2}. \label{eq:HEF-coeff-func}
\end{equation}
For completeness, we provide the details of the derivation of this result in~\ref{sec:app-A}, which to our knowledge are not present in the literature.%

The function ${\cal C}_{gi}$ in the complete LLA includes~\cite{Catani:1990xk,Catani:1990eg,Collins:1991ty,Catani:1994sq} some $\mu_F$-dependent terms which will not be compensated by the fixed-order evolution of GPDs, see Sec.~2.3 of Ref.~\cite{Lansberg:2021vie}. 
Therefore, to stay consistent with GPD evolution, we truncate the LLA down to the {\it doubly-logarithmic approximation (DLA)} which resums only corrections proportional to $\left(\alpha_s(\mu_R) \ln(1/\rho) \ln(\mu_F^2/{\bf k}_\perp^2) \right)^n$ in ${\cal C}_{gi}$, and has the following Mellin representation in the gluon channel~\cite{Collins:1991ty,Blumlein:1995eu}: 
\begin{equation}
    {\cal C}^{\rm (DL)}_{gg}(\rho,{\bf k}_\perp^2)= \int\frac{dN}{2\pi i}\, \rho^{-N}  \frac{\gamma_N}{{\bf k}_\perp^2}\left(\frac{{\bf k}_\perp^2}{\mu_F^2} \right)^{\gamma_N}, \label{eq:C-DLA-Mellin}
\end{equation}
with $\gamma_N={\hat{\alpha}_s}(\mu_R)/{N}$  with $\hat{\alpha}_s=\alpha_s C_A/\pi$. %
The Mellin transform in \ce{eq:C-DLA-Mellin} maps the logarithms of $1/\rho$ to poles at $N=0$: $\ln^{k-1}\frac{1}{\rho} \leftrightarrow \frac{(k-1)!}{N^k}$.

Substituting \ce{eq:C-DLA-Mellin} into \ce{eq:Master-formula-HEF}, one obtains the following Mellin-space result for the resummed coefficient function in the DLA HEF:
\begin{equation}
   C^{\text{(DLA HEF)}}_{g}(N)= \frac{-i\pi c}{2} \left(\frac{m_{Q}^2}{\mu_F^2}\right)^{\gamma_N} \frac{\pi\gamma_N}{\sin (\pi \gamma_N)}. \label{eq:CF-resumm-Mellin}
\end{equation}
which, expanded in $\alpha_s$ up to NNLO, %
reads in $\rho$ space (with an overall factor $-i\pi c/2$):
\begin{eqnarray}
  &\hspace{-7mm}\delta(|\rho|-1) + \frac{\hat{\alpha}_s}{|\rho|} \ln\left(\frac{m_{Q}^2}{\mu_F^2} \right)  +\frac{\hat{\alpha}_s^2}{|\rho|} \ln\frac{1}{|\rho|} \left[ \frac{\pi^2}{6} + \frac{1}{2}\ln^2\left(\frac{m_{Q}^2}{\mu_F^2}  \right)  \right].%
  \label{eq:resumm-C-expanded}
\end{eqnarray}
The  $\alpha_s^0$ and  $\alpha_s^1$ terms respectively agree with the imaginary part of $C_{g}^{\text{(0)}}$ and $C_{g}^{\text{(1, asy.)}}$. Using \ce{eq:C-gq}, one obtains a similar correspondence for $C_{q}^{\text{(1, asy.)}}$. Up to NNLO in $\alpha_s$, the DLA coincides with the complete LLA so the prediction of the NNLO term in the coefficient function in \ce{eq:resumm-C-expanded} is exact. 

The inverse Mellin transform of \ce{eq:CF-resumm-Mellin}, after subtraction\footnote{The subtraction is indicated by the $\check{~}$ mark on $C$.} of the  $\alpha_s^0$ term, can be cast into the following all-order expression in $\rho$ space, for $L_\mu\equiv\ln [m_{Q}^2/\mu_F^2]>0$:\footnote{$I_n$ ($J_n$) is the Bessel function of the second (first) kind and $\text{Li}_n(x)$ is the classic polylogarithm of 
order $n$.}
\begin{align}
  & \check{C}^{\text{(HEF)}}_{g}(\rho)=  \frac{-i\pi c}{2} \frac{\hat{\alpha}_s}{|\rho|} \sqrt{\frac{L_\mu}{L_\rho}} \times \label{eq:Analyt-resumm}
    \\ &\Biggl\{ I_1\left( 2\sqrt{L_\rho L_\mu} \right)  
    -2\sum\limits_{k=1}^\infty \text{Li}_{2k}(-1) \left(\frac{L_\rho}{L_\mu}\right)^{k} I_{2k-1}\left( 2\sqrt{L_\rho L_\mu} \right)  \Biggr\},\nn 
\end{align}
 where  $L_\rho\equiv\hat{\alpha}_s\ln 1/|\rho|$. 
  For $L_\mu<0$, $\check{C}^{\text{(HEF)}}_{g}(\rho)$  is obtained form the replacement $\sqrt{{L_\mu}/{L_\rho}} I_{2k-1}(2\sqrt{L_\mu L_\rho}) \to (-1)^{k} \sqrt{{(-L_\mu)}/{L_\rho}}J_{2k-1}(2\sqrt{-L_\mu L_\rho})$. %
 In addition, one has $\check{C}^{\text{(HEF)}}_{q}(\rho) = \frac{2C_F}{C_A} \check{C}^{\text{(HEF)}}_{g}(\rho)$. %

The low-$\rho$ behaviour of $\check{C}^{\text{(HEF)}}_{g}$ is governed by the rightmost singularity of \ce{eq:CF-resumm-Mellin} in the $N$ plane, namely at $N=\hat{\alpha}_s$. At $|\rho|\ll 1$, one thus has $\check{C}^{\text{(HEF)}}_{g} \propto |\rho|^{-\hat{\alpha}_s-1}$. This type of {\it hard Pomeron} behaviour  is absent in the case of the scale-fixing solution to the instability problem.
In a complete LLA treatment~\cite{BFKL1,BFKL2,BFKL3},
the value of the exponent  would be $4\hat{\alpha}_s\ln 2$ instead $\hat{\alpha}_s$. Anticipating the comparison with the experimental data, we note that the larger LLA value of the Pomeron intercept would lead to an even more rapidly growing cross section, incompatible with the measurements. 

\subsection{Matching of the NLO and resummed CF results}

So far, we have described two approximations for the CF coefficient function: the NLO CF of Sec.~\ref{sec:C-NLO}. %
and the DLA HEF %
of Sec.~\ref{sec:HEF-coeff-func}. %
Now we are in a position to combine them using a simple {\it subtractive-matching} prescription: 
\begin{eqnarray}
&& C_{g,q}^{\text{(match.)}}(x,\xi)=C_{ g,q}^{\text{(0)}}(x,\xi)+C_{ g,q}^{\text{(1)}}(x,\xi) \nonumber \\
&&+[\check{C}_{ g,q}^{\text{(HEF)}}\left( {\xi}/{|x|}\right) - C_{ g,q}^{\text{(1, asy.)}}(x,\xi) ] \theta(|x|-\xi). \label{eq:additive-matching}
\end{eqnarray}
For $\rho\to 0$ corresponding to $\xi \ll |x|< 1$, $C^{\text{(1, asy.)}}_{g,q}(x,\xi)\simeq C_{g,q}^{\text{(1)}}(x,\xi)$ by definition.  For $\rho\to 1$ corresponding to $\xi \lesssim x < 1$,   $\check{C}_{g,q}^{\text{(HEF)}}(\xi/|x|) \simeq C_{g,q}^{\text{(1, asy.)}}(x,\xi)$ because all terms proportional to $\ln^n (1/|\rho|)$ in \ce{eq:resumm-C-expanded} tend to 0.

In both limits, the matched result is given by the approximation we trust the most: ${C}_{ g,q}^{\text{(HEF)}}$ from DLA HEF for $\rho\to 0$ and $C_{ g,q}^{\text{(0)}}+C_{ g,q}^{\text{(1)}}$ from NLO CF for $\rho\to 1$. The $\theta$ function in \ce{eq:additive-matching} indicates that the resummation is applicable only for $|x|>\xi$, namely the DGLAP region.

One of many other possible HEF-to-CF-matching prescriptions is the {\it inverse-error-weighting (InEW)} matching~\cite{Echevarria:2018qyi,Lansberg:2021vie,Lansberg:2023kzf}. %
 Since our calculations in the inclusive case indicated~\cite{Lansberg:2021vie,Lansberg:2023kzf} that the central values and scale variation bands of the cross sections obtained with the subtractive and InEW matchings were consistent within the InEW matching uncertainty, we chose to stick here to the simplest  matching prescription, leaving the implementation of the InEW matching for a future work.     

\subsection{Resummed results}

As we have done to illustrate the known problematic NLO CF behaviour~\cite{Ivanov:2004vd}, let us first  examine  the relative $\mu_F$ uncertainty of the  NLO CF $\oplus$ DLA HEF results for the $t$-differential cross section shown in~\cf{fig:muF-ratio} in red: its $\mu_F$ uncertainty remains roughly constant and below a factor of 3, unlike the LO (grey) and NLO (blue) ones. This illustrates the improvement through the matched HEF resummation.

 \begin{figure}[!hbt]
    \centering
                 \captionsetup[subfloat]{captionskip=-0.5cm,margin={0.9\columnwidth,0cm}}
   \subfloat[]{ \includegraphics[width=\columnwidth]{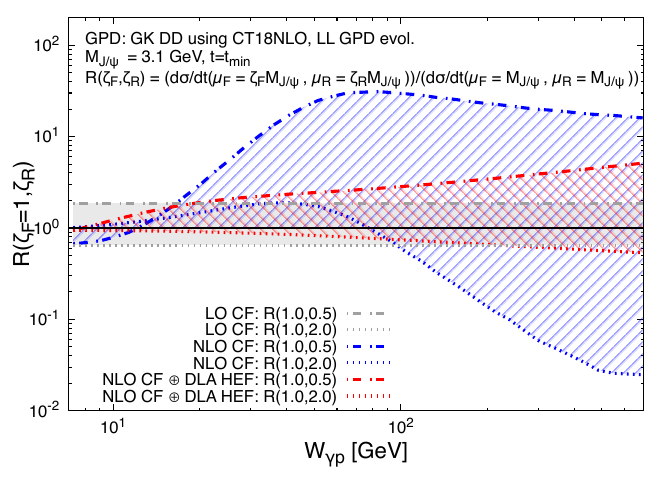}\label{fig:muR-ratio-jpsi}}\\\vspace*{-.5cm}
    \subfloat[]{ \includegraphics[width=\columnwidth]{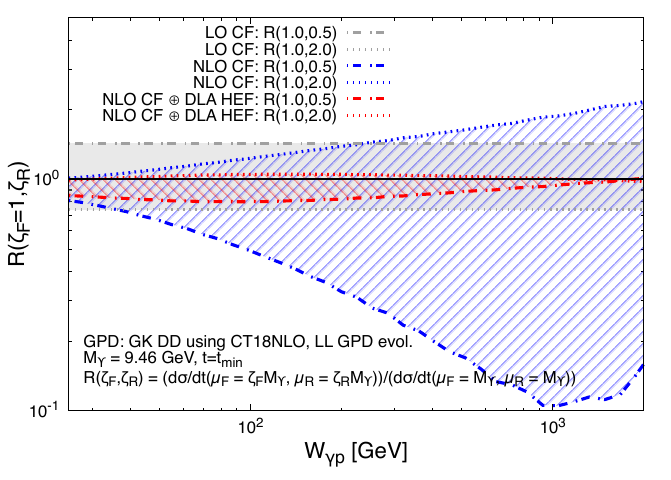}\label{fig:muR-ratio-upsilon}}\vspace*{-.2cm}
    \caption{ $\mu_R$ scale uncertainty of the LO CF, NLO CF and NLO CF $\oplus$ DLA HEF (a) $J/\psi$ and (b) $\Upsilon$ $t$-differential cross sections.
    \label{fig:muR-scale_var}}
\end{figure}

 Let us now turn to the $\mu_R$ variation/uncertainties of NLO CF $\oplus$ DLA HEF compared to the expected energy-independent LO CF variation, and that of the NLO CF shown in \cf{fig:muR-scale_var}.

 For the $J/\psi$ case (\cf{fig:muR-ratio-jpsi}), the NLO CF is again at odds with any expectations above $W_{\gamma p}\simeq20$~GeV. The NLO CF $\oplus$ DLA HEF uncertainty is significantly smaller than the NLO CF one and slowly increases to become larger than the LO CF one above $W_{\gamma p}\simeq 20$~GeV. This increase follows from the the $\mu_R$ dependence of the hard-pomeron contribution, whose asymptotic behaviour in the DLA (\ce{eq:Analyt-resumm}) directly depends on $\mu_R$ as $|\rho|^{-\hat{\alpha}_s(\mu_R)-1}$ without any compensation at this order.  We expect a smaller sensitivity with a complete NLLA computation.
 For the $\Upsilon$ case (\cf{fig:muR-ratio-upsilon}), the NLO CF band is consistently broader %
 than the LO CF one across the entire energy range shown and this broadening increases steadily with energy. Contrary to the $J/\psi$ case, the NLO CF $\oplus$ DLA HEF uncertainty does not increase and remains always (much) smaller than for the LO CF case. The aforementioned increase due to the hard-pomeron contribution is expected to set in at higher energies, even  above what is accessible at a possible FCC-eh.

\begin{figure}[!hbt]
    \centering
      \captionsetup[subfloat]{captionskip=-0.5cm,margin={0.9\columnwidth,0cm}}
        \subfloat[]{\includegraphics[width=\columnwidth]{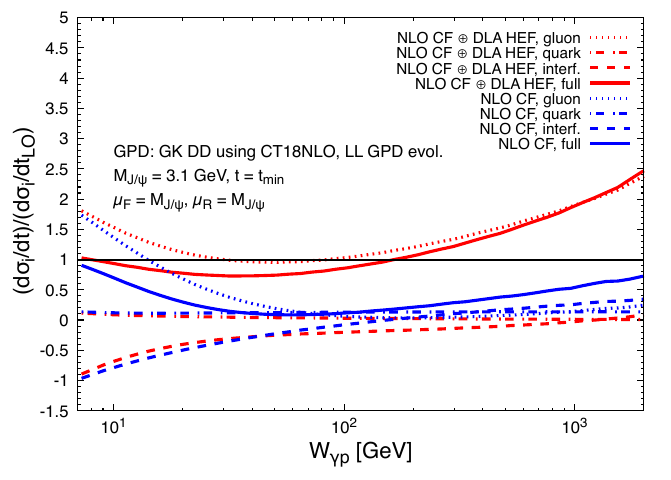}\label{fig:sigma_decomp-jpsi}}\vspace*{-.5cm}\\
          \subfloat[]{\includegraphics[width=\columnwidth]{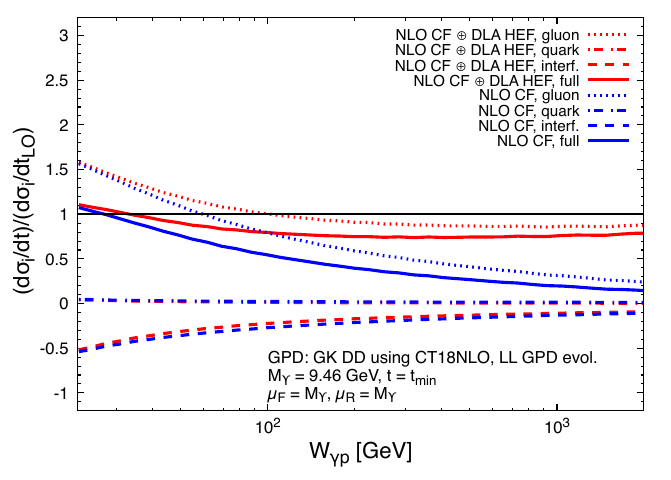}\label{fig:sigma_decomp-upsilon}}\vspace*{-.5cm}
    \caption{Decomposition of the NLO CF and NLO CF $\oplus$ DLA HEF of the $t$-differential cross section (normalised to the LO one) into contributions coming from only quarks, only gluons and their interference for (a) $J/\psi$  and (b) $\Upsilon$ .%
\label{fig:sigma_decomp}}
\end{figure}

In order to discuss the sensitivity of the cross section on the quark and gluon GPDs, one can decompose the cross section into the amplitude squared from quark-only (dash-dotted) and gluon-only (dotted) GPDs and their interference (dashed) as in \cf{fig:sigma_decomp}. For the NLO CF (blue), one observes that the full contribution (solid) is suppressed by the negative interference and is such that the gluon-only contribution becomes unnaturally small, even smaller than the quark-only contribution. This high-energy behaviour, noted in~\cite{Eskola:2022vpi,Eskola:2022vaf} is due to the instability of the NLO CF contributions and should not be considered as physically sound. Therefore, in what follows, we will not show the NLO CF cross sections anymore.  On the other hand, as also shown, the gluon-only contribution dominates the NLO CF $\oplus$ DLA HEF cross section, restoring the LO picture of gluon dominance and thus exhibiting an improved stability.%

\begin{figure}[!hb]
    \centering
      \captionsetup[subfloat]{captionskip=-0.5cm,margin={0.9\columnwidth,0cm}}
    \subfloat[]{\includegraphics[angle=-90,width=\columnwidth]{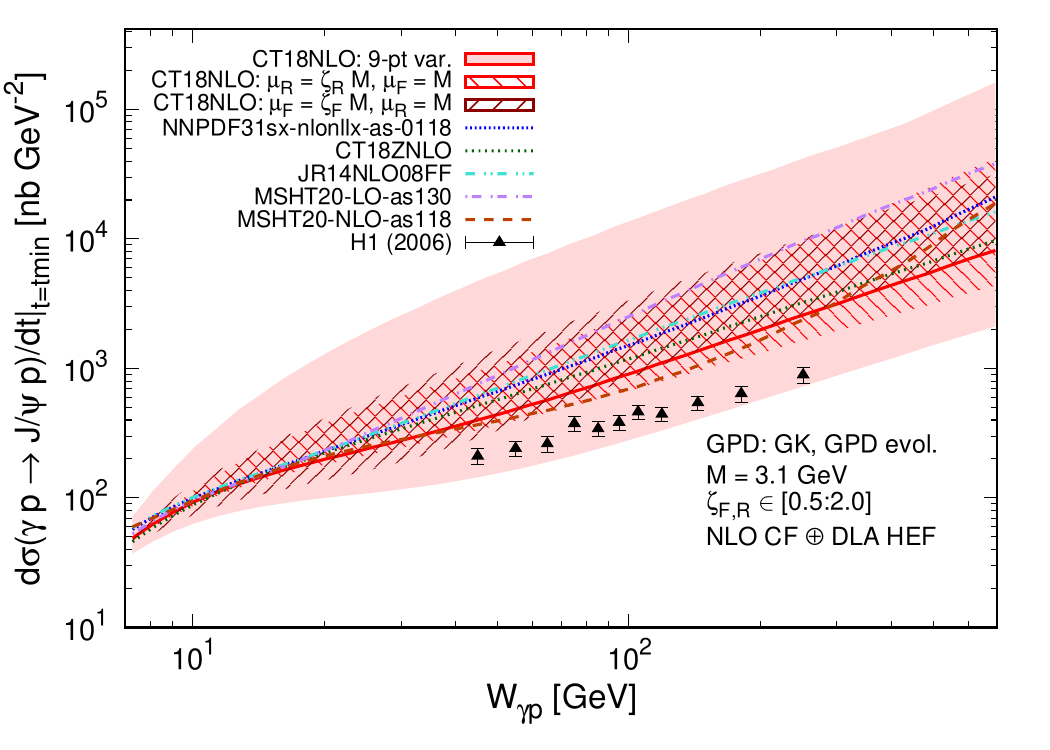}\label{fig:HEF_9pt-Jpsi}}\vspace*{-.5cm}\\
        \subfloat[]{\includegraphics[angle=-90,width=\columnwidth]{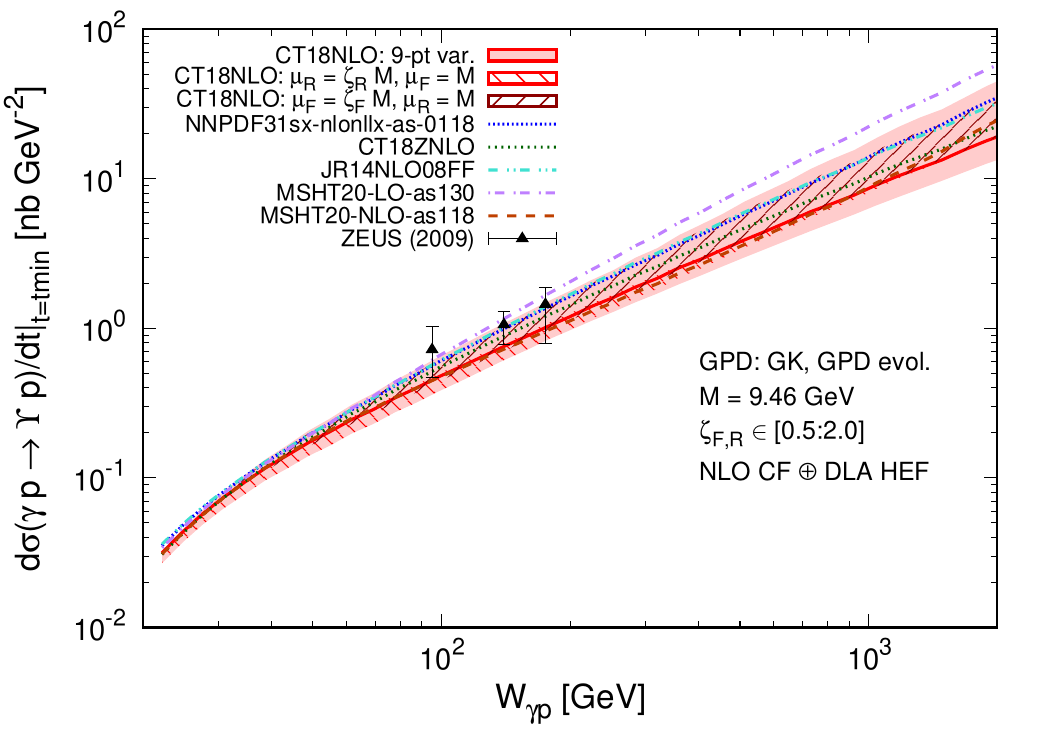}\label{fig:HEF_9pt-Upsi}}\vspace*{-.2cm}
    \caption{(a) NLO CF $\oplus$ DLA HEF cross section as a function of $W_{\gamma p}$ using a variety of input GPDs. The 9-point variation band is shaded while the hatched bands correspond to the sole variations of either 
    $\mu_F$ or $\mu_R$ with input CT18NLO GPDs.
    Also shown for comparison purposes is the most precise $t$-differential cross section data extrapolated down to $t=t_\text{min}$ from the H1 collaboration~\cite{H1:2005dtp}. (b) Same as (a) but for $\Upsilon$ production and $t$-differential cross section data extrapolated down to $t=t_\text{min}$ from the ZEUS collaboration~\cite{ZEUS:2009asc}. 
    \label{fig:HEF_9pt}
    }
\end{figure}

Having discussed the improvement brought about by the HEF resummation with regards to the stability of the cross section, we are now ready to compare our results to experimental data. As announced, we focus on the $t$-differential cross section at $t=t_\text{min}$. To make our comparison even more accurate, instead of simply comparing our computation to the data in the smallest $t$ bin, we have converted them to $t=t_\text{min}$
extrapolating the measured $t$ dependence at $\langle |t| \rangle = 0.03~{\rm GeV}^2$ as follows $d\sigma/dt \propto \exp(-b(W_{\gamma p})t)$ with $b(W_{\gamma p})\,{\rm GeV}^2= B_0 + 4\alpha_{{\mathbb P}}' \ln(W_{\gamma p}/90)$ and duly propagating the experimental uncertainties. We take $B_0 = 4.63$ and $\alpha_{{\mathbb P}}' = 0.164$~\cite{H1:2005dtp}. For the record, the obtained corrections is on the order of 15-20\%.
Our computed cross sections are shown with an estimated uncertainty following from the 9-point scale variation\footnote{$\mu_R$ and $\mu_F$ are independently set to $\mu_F=\mu_0\times {\zeta_F}$ and $\mu_R=\mu_0\times {\zeta_R}$, for $\zeta_{F,R} \in \{1/2,1,2\}$ and $\mu_0=M_\Q$. The filled band is the envelope of the corresponding cross sections. As a complement, we show the envelope corresponding to the sole variation of $\mu_F$ and $\mu_R$ with hatched bands.} about $M_\Q$.

In addition, we have added to~\cf{fig:HEF_9pt} results computed with the central eigenvector set of CT18ZNLO~\cite{Hou:2019efy}, an alternate CT18 fit with a different gluon PDF, as well as the LO set MSHT20-LO-as130~\cite{Bailey:2020ooq}, another conventional NLO set MSHT20-NLO-as118~\cite{Bailey:2020ooq}, as well as JR14NLO08FF~\cite{Jimenez-Delgado:2014twa} and NNPDF31sx-nlonllx-as-0118~\cite{Ball:2017otu}, both known to be steeper at low $x$ and low scales. It allows us to assess the uncertainty from the input PDF in our DD GPDs to be compared to the pertubative uncertainties evaluated from the scale variation.

 For the $J/\psi$ case, our (large) scale-uncertainty band shown in \cf{fig:HEF_9pt-Jpsi} is in agreement with the experimental data\footnote{We however stress that the size of the uncertainty is a matter of convention as we could vary the scale by a different amount than 2.} which tends to lie on the lower side.
 The coloured curves obtained for the central values of different PDF sets and with our central scale $M_\Q$ exhibit an uncertainty of ${\cal O}(2-3)$ which is similar to the uncertainty from varying $\mu_F$ and $\mu_F$ separately. For $\Upsilon$, shown in \cf{fig:HEF_9pt-Upsi}, our band is again in agreement with the experimental data, with a similar spread in the central values at the largest energies shown. In this case, the PDF-input uncertainty is similar than the combined scale uncertainty from the 9-point scale variation.

\section{Conclusions}

In the present paper we have discussed exclusive vector-quarkonium production in the framework of CF at NLO supplemented by the resummation of a series of higher-order corrections in CF, which are crucial to resolve the perturbative instability of the NLO CF computation at $W_{\gamma p} \gg M_{\Q}$. In our opinion, this approach is the most suitable in providing a uniformly accurate description of this observable over a wide energy range, in comparison to e.g. CGC~\cite{Mantysaari:2022kdm}, small-$x$~\cite{Boer:2023mip} or $k_T$-factorisation~\cite{Jones:2013pga} approaches, applicable only at high energies. 

We have found that the resummation indeed cures the NLO CF perturbative instability with a smooth energy dependence in agreement with the experimental data. Our results also confirm that quarkonium photoproduction is predominantly gluon-induced with quark-induced contributions only arising from radiative corrections and at most at the level of 20\% via interference with gluon-induced contributions for $W_{\gamma p} > 100$~GeV.

We think our study is an important milestone for future global fits of GPDs using quarkonium data. It also highlights that most of the observed features of the NLO CF and resummation results (e.g. scale and flavour dependencies)  cannot be captured by reducing GPDs to a simple square of PDFs. Constraints on PDFs might however be obtained from a global GPD fit with PDFs as forward inputs in a DD model for instance. Quantifying such sensitivity along with the expected experimental improvements at the LHC~\cite{Chapon:2020heu} or the EIC~\cite{Boer:2024ylx} is clearly beyond the scope of the present work. 

Our results are limited by the large scale uncertainty in the $J/\psi$ case (\cf{fig:HEF_9pt-Jpsi}), for which the dataset is the most extensive. The picture is more promising for $\Upsilon$ photoproduction (\cf{fig:HEF_9pt-Upsi}), although the dataset is more restricted. We anticipate the scale uncertainty will be reduced once we develop this computation beyond the DLA (see e.g. \cite{Nefedov:2024swu}), which is the subject of a future work. Further improvement can be made by including relativistic-$v^2$ and/or higher-twist corrections in $\Lambda_{{\rm QCD}}/M_\Q$, which should be larger for $J/\psi$ than for $\Upsilon$ production.

\paragraph*{\bf Acknowledgements.}
We thank V.~Bertone,  D.~Boer, V.~Braun, J.R. Cudell, H.~Dutrieux, K. Lynch,  R. McNulty, H. Moutarde, M.~Ozcelik, M. Ryskin, M. Strikman, L.~Szymanowski, C. Van Hulse, S. Wallon, F.~Yuan for useful discussions.
This project has received funding from the European Union's Horizon 2020 research and innovation programme under grant agreement No.~824093 in order to contribute to the EU Virtual Access {\sc NLOAccess} and the JRA Fixed-Target Experiments at the LHC and a Marie Sk{\l}odowska-Curie action ``RadCor4HEF'' under grant agreement No.~101065263.
This project has also received funding from the Agence Nationale de la Recherche (ANR) via the grant ANR-20-CE31-0015 (``PrecisOnium'') and via the IDEX Paris-Saclay ``Investissements d’Avenir'' (ANR-11-IDEX-0003-01) through the GLUODYNAMICS project funded by the ``P2IO LabEx (ANR-10-LABX-0038)''.
This work  was also partly supported by the French CNRS via the IN2P3 projects ``GLUE@NLO'' and ``QCDFactorisation@NLO'' as well as via the COPIN-IN2P3 project \#12-147 ``kT factorisation and quarkonium production in the LHC era''.

\vspace*{-0.25cm}
\appendix\vspace*{-0.25cm}
\renewcommand{\thefigure}{\arabic{figure}}

\section{derivation of the HEF coefficient function}\label{sec:app-A}

In this appendix we provide details of the derivation of the HEF coefficient function (\ref{eq:HEF-coeff-func}). Let us consider the one-loop amplitudes of the processes:
\begin{eqnarray}
     \gamma(q)+g(p_1)\to Q\bar{Q}\left[{}^3S_1^{[1]} \right](p_{\Q}) + g(p_1'), \label{proc:ga+g-cc+g} \\
    \gamma(q)+q(p_1)\to Q\bar{Q}\left[{}^3S_1^{[1]} \right](p_{\Q}) + q(p_1'), \label{proc:ga+q-cc+q}
\end{eqnarray}
 with typical diagrams shown in   \cf{fig:diags-1L-gq}(a,b). One is interested in the imaginary part of these amplitudes in the Regge limit, when the partonic centre-of-mass energy $\hat{s}=(q+p_1)^2$ is much larger than $\hat{t}=(q-p_{\Q})^2=t_{\min}$ and $M_{\Q}^2$. %
The Regge asymptotics can be computed with the help of the following replacement for the Feynman-gauge $t$-channel gluon propagators highlighted in bold in~\cf{fig:diags-1L-gq}:
\begin{equation}
   \frac{ -ig^{\mu\nu}}{k^2+i\varepsilon} \to \frac{i}{2 {\bf k}_\perp^2}(n_-^{\mu} n_+^{\nu}+n_+^{\mu} n_-^{\nu}),\label{eq:Gribov}
\end{equation}
where $n_{+}^\mu = 2q^\mu/\sqrt{\hat{s}}$ and $n_-^\mu = 2p_1^\mu/\sqrt{\hat{s}}$ are the Sudakov basis vectors\footnote{We define the Sudakov decomposition for any vector $k^\mu=(k^+n_-^\mu + k^-n_+^\mu)/2 + k_\perp^\mu$ with $n_{\pm}\cdot k_\perp=0$, $k^{\pm}=n_{\pm}\cdot k$ such that $k^2=k_+k_--{\bf k}_\perp^2$.} with $n^+.n^-=2$. %
After this replacement, the one-loop gluon ($i=g$, Eq.~(\ref{proc:ga+g-cc+g})) or quark ($i=q$, Eq.~(\ref{proc:ga+q-cc+q})) amplitude can be factorised as: %
\begin{equation}
    i{\cal M}_{\gamma i} = -\frac{1}{2}\int \frac{d^{2-2\epsilon}{\bf l}_\perp}{(2\pi)^{4-2\epsilon}} \int \frac{dl_+ dl_-}{2} \frac{{\cal A}_{ab}(l_+,{\bf l}_\perp) {\cal B}^{ab}_{i}(l_-,{\bf l}_\perp)}{(2{\bf l}_\perp^2)^2}, \label{eq:ampl_ga+i-cc+i}
\end{equation}
with the the projectile (target) Impact Factors~(IF) ${\cal A}_{ab}$ (${\cal B}^{ab}_i$) defined below. %
In the leading-power approximation w.r.t. $1/\hat{s}$, the $l^+$ integration gets routed through the projectile IF ${\cal A}_{ab}$, while the $l_-$ integration goes through the target IF ${\cal B}_i^{ab}$. The result for the projectile IF before the $l_+$ integration is:
\begin{equation}
{\cal A}_{ab}= \frac{-32c q_-^2 {\bf l}_T^2 (\varepsilon^{*\Q}. \varepsilon^{\gamma})}{[2l_+q_--M_{\Q}^2-4{\bf l}_\perp^2+i\varepsilon] [2l_+q_-+M_{\Q}^2+4{\bf l}_\perp^2-i\varepsilon]},  
\end{equation}
where $\varepsilon^{\gamma}(q). p_1=0$, $a,b$ are the colour indices of $t$-channel gluons and in the derivation we have used the following replacement for the spinors of outgoing heavy quark and antiquark with momenta $p_{\Q}/2$:
\begin{equation}
    u_\alpha^i\left(\frac{p_{\Q}}{2}\right) \bar{v}_\beta^j\left(\frac{p_{\Q}}{2}\right) \to  \delta^{ij} \frac{[(\slashed{p}_{\Q}-M_{\Q}) \slashed{\varepsilon}^{*(\Q)} (\slashed{p}_{\Q}+M_{\Q})]_{\alpha\beta}}{4 \sqrt{M_{\Q}^3 N_c } } ,
\end{equation}
to project their colour($i,j$) and Dirac ($\alpha,\beta$) %
 indices on the colour-singlet state with the total spin equal to one. Integrating out the $l_+$ momentum in the projectile IF gives us the HEF coefficient function (\ref{eq:HEF-coeff-func}) as expected:
\begin{equation}
\int\limits_{-\infty}^{+\infty} dl_+ {\cal A}_{ab}(l_+,{\bf l}_\perp) = 8 i \delta_{ab} (\varepsilon^{*\Q}.\varepsilon^{\gamma}) \frac{q_- {{\bf l}_\perp^2}}{m_{Q}^2} h({\bf l}_\perp^2). \label{eq:proj-IF}\\    
\end{equation}

\begin{figure}[hbt!]
    \centering
    \includegraphics[width=0.35\textwidth]{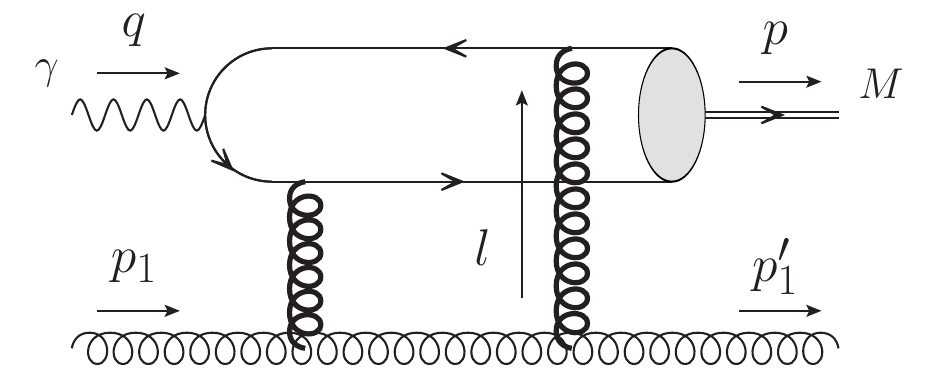}
    \\
    (a)
    \\[5pt]
    \includegraphics[width=0.35\textwidth]{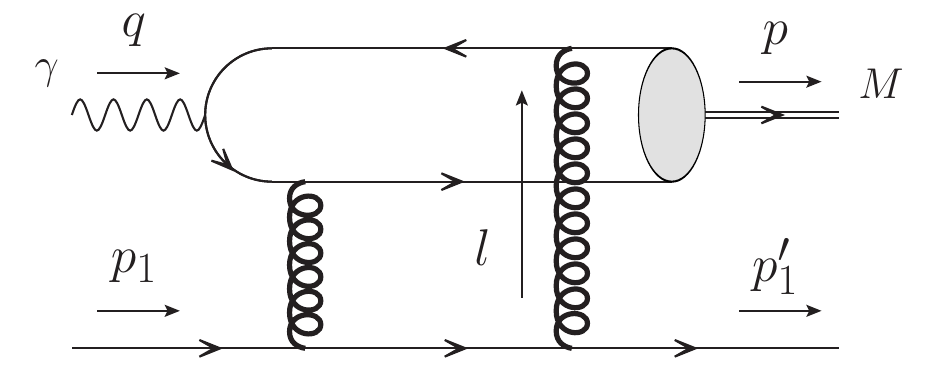}
    \\
    (b)
    \caption{Typical Feynman diagrams contributing to the imaginary part of the amplitude in the gluon-target (a) and quark-target (b) case. All $6$ possible diagrams should be included.}
    \label{fig:diags-1L-gq}
\end{figure}

To project the gluon-target IF on the CF coefficient function in  Eq.~(\ref{eq:Ampl-CF}), we replace the polarisation vectors of the target gluons as follows:
\begin{equation}
    \varepsilon_\mu(p_1) \varepsilon^*_\nu(p_1') \to  \frac{\delta_{ab}}{(N_c^2-1)} \frac{ -g_{\mu\nu} + \left(n_+^\mu n_-^\nu + n_+^\nu n_-^\mu \right)/2 }{2(D-2)} ,
\end{equation}
with $D=4-2\epsilon$ being the spacetime dimension, while for target quarks the corresponding replacement for spinors is:
\begin{equation}
     u_\alpha^i(p'_1) \bar{u}_\beta^j(p_1) \to \frac{\delta_{ij}}{N_c} \frac{(\slashed{p}_1)_{\alpha \beta}}{2x}.
\end{equation}
In the CF kinematics, $p_1^\mu=x P^\mu$ with $P$ being the proton momentum and $\xi\ll |x|\lesssim 1$ in the $|\rho|\ll 1$ region we are interested in. %
With these substitutions, the result for the target IFs before the $l_-$ integration are:
\begin{eqnarray}
 && \hspace{-13mm}  {\cal B}^{ab}_{g} = - \frac{4N_c\delta^{ab} g_s^2}{N_c^2-1} \frac{(p_1^+)^2 {\bf l}_\perp^2}{(l_- p_1^++{\bf l}_\perp^2-i\varepsilon)(l_- p_1^+-{\bf l}_\perp^2+i\varepsilon)}, \\
 && \hspace{-13mm} {\cal B}^{ab}_{q}(l_-,{\bf l}_\perp) = \frac{2C_F}{xC_A}  {\cal B}^{ab}_{g}(l_-,{\bf l}_\perp), \label{eq:Bq-before-int}  %
\end{eqnarray}
while the gluon IF after integrating-out $l_-$ gives:
\begin{equation}
\int\limits_{-\infty}^{\infty} dl_- {\cal B}_{g}^{ab}(l_-,{\bf l}_\perp) =  4\pi i g_s^2\frac{N_c\delta^{ab}}{N_c^2-1}  |p_1^+|. \label{eq:target-IF-gg}
\end{equation}
 Substituting these results into Eq.~(\ref{eq:ampl_ga+i-cc+i}), one obtains the imaginary part of the NLO CF gluon coefficient function in the Regge limit $|\rho|\ll 1$ in the following form:
\begin{equation}
    \text{Im}\, C_{g}^{\text{(NLO)}}(\rho\ll 1) = - \frac{r_\Gamma}{|\rho|} \int\limits_0^{\infty} d{\bf l}_\perp^2 \left(\frac{\mu^2}{{\bf l}_\perp^2} \right)^\epsilon \left\{ \frac{\hat{\alpha}_s}{{\bf l}_\perp^2} \right\} h({\bf l}_\perp^2), \label{eq:ImCg-NLO-lT}
\end{equation}
where $r_\Gamma=(4\pi)^\epsilon/\Gamma(1-\epsilon)$. On the one hand, integrating out the ${\bf l}_\perp^2$ in the last equation and subtracting the collinear pole according to the $\overline{{\rm MS}}$ scheme, one recovers the asymptotic result (\ref{eq:Cg-NLO:low-X}). On the other hand, Eq.~(\ref{eq:ImCg-NLO-lT}) at $\epsilon=0$ has the form of Eq.~(\ref{eq:Master-formula-HEF}) with the factor in the curly brackets in  Eq.~(\ref{eq:ImCg-NLO-lT}) being nothing but the $O(\alpha_s)$ term in the expansion of the resummation factor (\ref{eq:C-DLA-Mellin}). The last observation fixes the overall factor in  Eq.~(\ref{eq:Master-formula-HEF}) for both the quark and gluon cases. %

 Finally, we give a hint for the derivation of  Eq.~(\ref{eq:Analyt-resumm}). The Mellin transform of Eq.~(\ref{eq:CF-resumm-Mellin}) can be rewritten as an integral over $\gamma=\hat{\alpha}_s/N$:
 \begin{equation*}
          C^{\text{(HEF)}}_{g}(\rho)= \frac{-i\pi c}{ 2}\frac{\hat{\alpha}_s}{|\rho|}\sum\limits_{n=0}^\infty \frac{L_\rho^n}{n!} \oint \frac{d\gamma}{2\pi i} \gamma^{-n-2} e^{\gamma L_\mu} \frac{\pi\gamma}{\sin(\pi\gamma)},
 \end{equation*}
 where we have used the series expansion of $e^{L_\rho/\gamma}$. The contour in the $\gamma$ plane encircles the singularity at $\gamma=0$ but can be blown-up to encircle poles at positive and negative integer points instead which, after summing all the residues, gives $(-1)^{n+1} \left[ \text{Li}_{n+1} \left(-e^{-L_{\mu}}\right) + (-1)^{n+1}\text{Li}_{n+1} \left(-e^{L_{\mu}}\right)  \right]$ for the $\gamma$ integral. This combination of polylogarithms can be rewritten in terms of logarithms using the identity~\cite{DilogID}:
\[
\text{Li}_{n}(z) + (-1)^n \text{Li}_{n}\left(\frac{1}{z}\right) = -\frac{\ln^n(-z)}{n!} + 2 \sum\limits_{k=1}^{\lfloor \frac{n}{2} \rfloor} \frac{\text{Li}_{2k}(-1)}{(n-2k)!} \ln^{n-2k}(-z), 
\]
which after swapping the order of summations over $n$ and $k$ and summing the infinite series in $n$ into Bessel functions, gives Eq.~(\ref{eq:Analyt-resumm}).

\vspace*{-0.5cm}

\bibliographystyle{utphys}
\biboptions{sort&compress}
\bibliography{NLO-Exclusive-Photoproduction}

\end{document}